\documentclass[sigconf]{acmart}

\usepackage{multirow}
\usepackage{xcolor}
\usepackage{array}

\AtBeginDocument{%
  }

\copyrightyear{2024}
\acmYear{2024}
\setcopyright{rightsretained}
\acmConference[UIST '24]{The 37th Annual ACM Symposium on User Interface Software and Technology}{October 13--16, 2024}{Pittsburgh, PA, USA}
\acmBooktitle{The 37th Annual ACM Symposium on User Interface Software and Technology (UIST '24), October 13--16, 2024, Pittsburgh, PA, USA}
\acmDOI{10.1145/3654777.3676413}
\acmISBN{979-8-4007-0628-8/24/10}

\definecolor{darkgreen}{rgb}{0.0, 0.5, 0.0}

\begin{document}

\title[EVE: Enabling Anyone to Train Robots using Augmented Reality]{\includegraphics[height=5mm]{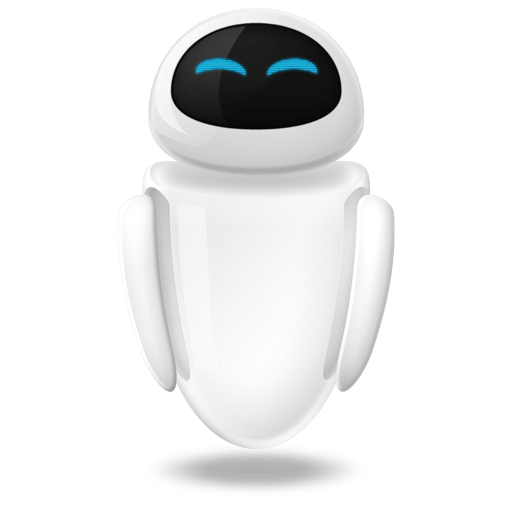}EVE: Enabling Anyone to Train Robots using Augmented Reality}

\author{Jun Wang}
\affiliation{
    \institution{University of Washington}
    \country{junw3@cs.washington.edu}
}

\author{Chun-Cheng Chang$^*$}
\affiliation{
    \institution{University of Washington}
    \country{chuncc@cs.washington.edu}
}

\author{Jiafei Duan$^*$}
\affiliation{
    \institution{University of Washington}
    \country{duanj1@cs.washington.edu}
}

\author{Dieter Fox}
\affiliation{
    \institution{University of Washington}
    \institution{NVIDIA}
    \country{fox@cs.washington.edu}
}

\author{Ranjay Krishna}
\affiliation{
    \institution{University of Washington}
    \institution{Allen Institute for AI}
    \country{ranjay@cs.washington.edu}
}

\thanks{$^*$Equal Contribution.}
\renewcommand{\shortauthors}{Wang et al.}

\begin{abstract}
The increasing affordability of robot hardware is accelerating the integration of robots into everyday activities. However, training a robot to automate a task requires expensive trajectory data where a trained human annotator moves a physical robot to train it. 
Consequently, only those with access to robots produce demonstrations to train robots. 
In this work, we remove this restriction with EVE, an iOS app that enables everyday users to train robots using intuitive augmented reality visualizations, without needing a physical robot. 
With EVE, users can collect demonstrations by specifying waypoints with their hands, visually inspecting the environment for obstacles, modifying existing waypoints, and verifying collected trajectories. 
In a user study ($N=14$, $D=30$) consisting of three common tabletop tasks, EVE outperformed three state-of-the-art interfaces in success rate and was comparable to kinesthetic teaching—physically moving a physical robot—in completion time, usability, motion intent communication, enjoyment, and preference ($mean_{p}=0.30$). 
EVE allows users to train robots for personalized tasks, such as sorting desk supplies, organizing ingredients, or setting up board games. We conclude by enumerating limitations and design considerations for future AR-based demonstration collection systems for robotics.

\end{abstract}



\maketitle
\section{Introduction}
\begin{figure}
    \centering
    \includegraphics[width=\linewidth]{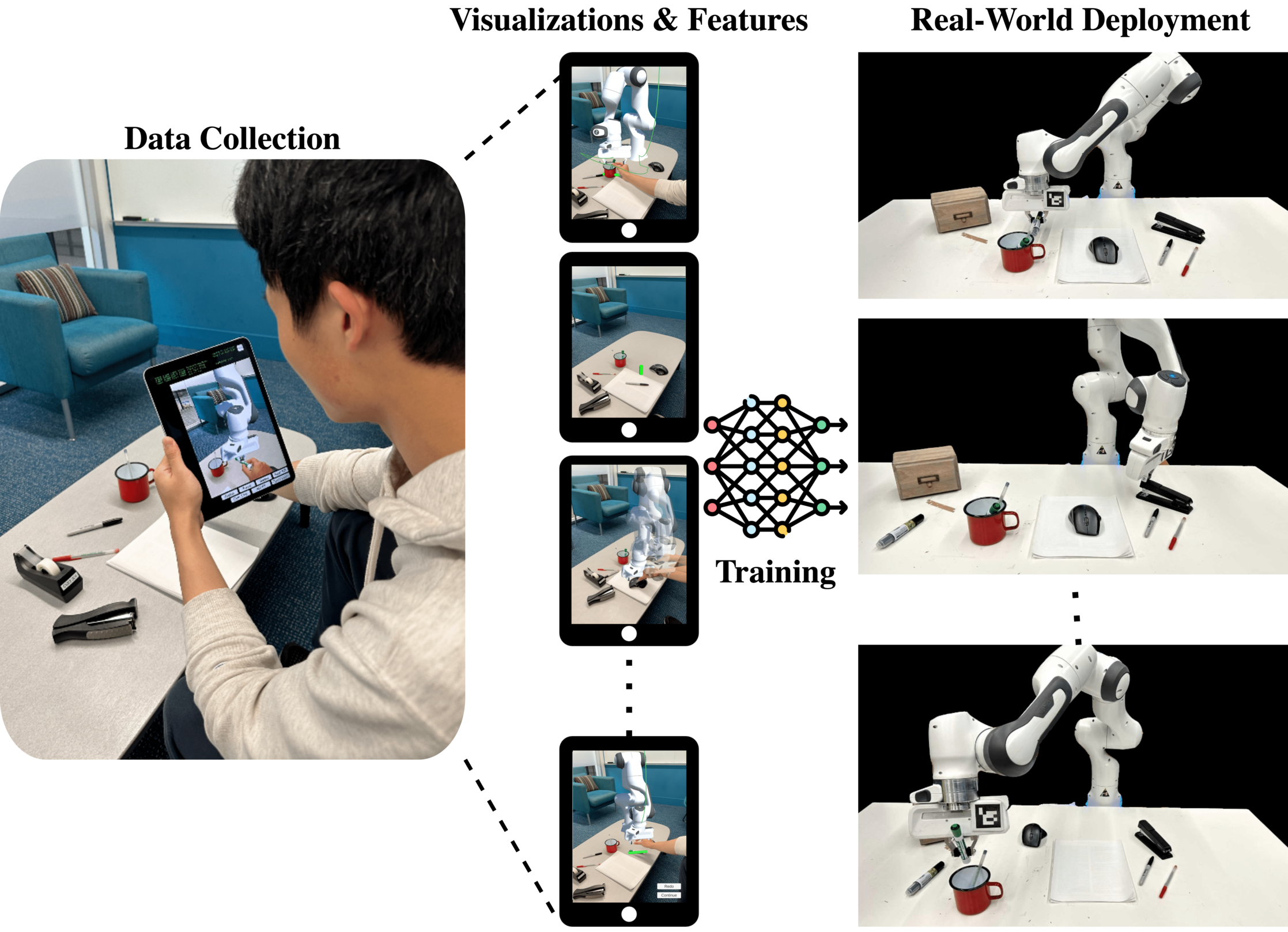}
    \caption{EVE allows everyday users to collect data to train a real robot using intuitive augmented reality (AR) visualizations without needing a physical robot. Our application enables users to move the AR robot by setting waypoints with hand gestures, visually inspect the real-world environment to avoid obstacles, modify the collected trajectory, and verify the data collection by replaying the task with the AR robot. Videos of AR visualizations, features, and real-world deployment are available at \href{https://junwang0510.github.io/EVE/}{https://junwang0510.github.io/EVE/}.}
    \label{fig:teaser}
    \Description{An example collection with EVE. A person is holding an iPad with one hand and using the other to rearrange items on the table (left). Four examples of AR visualizations that assist the user in gathering data are shown (middle left). A neural network diagram represents using the collected data to train a real robot (middle right). Three images demonstrate the real-world deployment (right).}
\end{figure}

The decreasing costs of robot components are accelerating the integration of robots into everyday life. With companies offering robotic arms at a few hundred dollars, portable, lightweight, and mobile robots are being trained to assist individuals with a wide range of everyday tasks~\cite{mechArmPi2023,HelloRobot2023,shaw2023leap}. However, training a robot to automate a task often requires expensive manually curated human trajectories. Annotators are hired and trained to kinesthetically move or teleoperate a physical robot. The robot's joints and gripper are moved into a sequence of configurations to accomplish a task. For example, to train a robot to clean a table, annotators move the gripper towards a cloth, close the gripper to grasp the cloth, and then begin moving it around to wipe the table.

There are no available systems for everyday users to train robots without first purchasing an expensive robot. Users who are considering purchasing a robot to automate a task in their homes need to first purchase the robot. Next, they need to learn how to move bulky machinery to produce useful trajectories. Consequently, only a limited number of roboticists and developers with access to robots will determine the range of consumer robotic applications.

Recently, a data collection mechanism, called AR2-D2, demonstrated the possibility of training robots on simple tasks without needing a physical robot~\cite{duan2023ar2}. AR2-D2 converts videos of people manipulating objects in their homes into robot trajectories. AR2-D2 uses computer vision models to automatically capture the 3D structure of the environment for training robot policies. AR2-D2 demonstrated that the quality of data gathered using videos is comparable with data collected using traditional kinesthetic or teleoperated approaches with real robots. However, AR2-D2 was only evaluated on simple, short-horizon tasks such as pressing a computer mouse or picking up a plastic bowl. Ecologically valid user tasks, such as cleaning a table, require multiple interactions between the user and their environment, resulting in longer videos. Since the tasks are longer, users are more likely to take wrong actions and need to backtrack and redo particular steps. Thus, while AR2-D2 shows that training a robot might not require a real robot, there is a need to explore the possible designs that make such a system usable and ecologically valid from a user-centric perspective.

We introduce EVE, an iOS app that allows everyday users to train robots using an augmented reality (AR) robot, without requiring a physical robot. To train a real robot, users use EVE to project an AR robot in their environment and guide its behavior. Users can specify waypoints that define the trajectory the AR robot should follow. During collection, users can visually inspect the movement of the AR robot from different angles to ensure that the AR robot avoids obstacles. Users can modify the collected trajectory by reverting the AR robot's position to a previous waypoint. EVE also contains a number of visualization features that provide users with confidence that their trajectory could be used to train a real robot.

To understand the opportunities and challenges of AR-based trajectory collection, we conduct a formative study with $10$ participants with varying experience in robotics, comparing AR2-D2 to the initial prototype of EVE with three different AR visualizations. From this study, we identify six usability challenges users faced and implemented seven additional features to improve EVE's accuracy, flexibility, and verifiability.

We design a user study ($N=14$) to quantitatively compare EVE against state-of-the-art trajectory collection interfaces on three tabletop tasks. EVE achieved the highest task success rate and performed comparably to kinesthetic teaching in task completion time, system usability, motion intent communication, user enjoyment, and overall preference ($mean_{p}=0.30$). Participants customized their workflows with multiple visualizations and features to collect demonstrations for complex, long-horizon tasks.
Additionally, we train machine learning policies with the collected trajectories and deploy the policies on a real robot for the toggle switch task. Specifically, we collect $6$ trajectories using both AR2-D2 and EVE and train policies for $30,000$ iterations. We find that on $30$ evaluations, policies trained with EVE resulted in $10\%$ higher accuracy.

EVE facilitates personalized robot training in everyday environments, allowing users to collect demonstrations tailored to their specific needs and adapt to various robot platforms and motion planners. There are still numerous unanswered questions: the current system lacks support for bimanual manipulation and mobile robot data collection, which are crucial for many real-world tasks like washing dishes and cleaning floors. Future research should investigate the use of AR head-mounted displays for bimanual control and the integration of enhanced depth cues for more precise manipulation. Moreover, incorporating Simultaneous Localization and Mapping (SLAM) techniques with EVE could enable multi-view data collection, which is essential for mobile robots. Enhancements in task variability are also necessary, such as through more efficient motion planners and hardware improvements, to ensure robust and accurate performance in dynamic and unstructured settings.

\section{Related work}
As we aim to enhance the accuracy, usability, and verifiability of an AR-based trajectory collection interface for robotics, our work builds upon prior research on human trajectory collection mechanisms and the use of AR to assist in motion intent communication.

\subsection{Trajectory collection for robotics}
A common approach for collecting demonstration data includes kinesthetic teaching, where an annotator physically guides the robot through a desired trajectory~\cite{KTSurvey}. Despite the direct haptic feedback, this mechanism can be slow and tedious~\cite{osentoski2010crowdsourcing} as users have to physically move a bulky, heavy machine by rotating its joints into desired configurations.

Another popular collection method is teleoperation, where the user controls the robot remotely using specialized controllers, such as 3D space mouse~\cite{zhu2022viola}, game or virtual reality (VR) controllers~\cite{ebert2021bridge,brohan2022rt,seo2023deep,Mimic}, smartphones~\cite{mandlekar2018roboturk}, and haptic devices~\cite{shaw2023videodex,toedtheide2023force}. These specialized devices are costly and challenging to use due to high latency and unintuitive controls for modifying multiple robot joints~\cite{Akgn2011RobotLF}.

More recent interfaces capture human expert demonstrations with physical robots using specialized hardware like ALOHA~\cite{zhao2023learning}, GELLO~\cite{wu2023gello}, DexCap~\cite{wang2024dexcap}, and UMI~\cite{chi2024universal}. These hardware resemble gloves in the shape of the robot gripper, offering users greater flexibility in manipulating objects with the robot without needing to control the entire robot arm. However, these studies often prioritize data collection \textit{efficiency} and robot \textit{performance}, with minimal user studies examining user preference and usability~\cite{kapadia2017echobot,szafir2021connecting}.

Prior work compares kinesthetic teaching and teleoperation on simple tasks such as manipulating cups~\cite{comp, Akgn2011RobotLF}. The common finding is that kinesthetic teaching is preferred over teleoperation based on collection time, ease of use, and accuracy of demonstrations. By contrast, our paper proposes using an iOS app with AR to train robots. We focus on three realistic tabletop tasks with varying levels of difficulty, leading to a more comprehensive comparison between different collection interfaces.

Most closely related to our work is AR2-D2~\cite{duan2023ar2}, a data collection mechanism that does not require real robots or trained human annotators, and resituates trajectory collection outside a robotics lab with AR. AR2-D2 shows that users require no training in contrast to common demonstration collection interfaces. However, AR2-D2's utility is limited; it focuses on simple, short-horizon tasks like clicking a button. They also do not explore the design space for developing a prototype for everyday users.

Our system, EVE, focuses on the human-centric aspect of the trajectory collection process with new visualizations and new features derived from a formative study. It allows users to select suitable collection methods based on different ecologically valid tasks. Furthermore, we examine the system's accuracy, speed, usability, and verifiability through three common, long-horizon tabletop tasks.

\subsection{Augmented reality and robotics}
Traditional approaches to improving human-robot interaction, such as through robots' movements~\cite{Dragan2013, PlanningSA}, gestural motion~\cite{ModelingAE, hri2012}, and gaze outputs~\cite{Mutlu2006ASR, hri2018}, are limited due to the challenge of presenting expressive physical feedback that goes beyond internal capabilities. By contrast, AR provides visual feedback in one's line of sight, opening up exciting new opportunities for human-robot interaction research. A recent taxonomy of AR and robotics has identified five purposes and benefits of visual augmentation~\cite{Suzuki2022ARRobotics}. We focus on analyzing work related to motion intent communication since our goal is to improve system usability by allowing users to easily control and understand the goal of the AR robot.

AR interfaces help communicate the robot's intention to the user through spatial information. Prior work has experimented with different designs based on factors such as path, waypoints, velocity, arrival times, and shadows~\cite{Pascher_2023,communicateAR,flyar,PinpointFly}. 
Most relevant to our work is Rosen et al.~\cite{rosen2017}, who visualized the robotic arm's motion intent by overlaying planned robot poses with a mixed reality head-mounted display. Evaluating against 2D displays and no visualizations, they found that the AR visualization allows users to more quickly ($-62\%$ in time) and accurately ($+16\%$ in accuracy) identify where the robot is going to move. They also iterated on the system to allow the robot to replan an intended trajectory with MoveIt motion planner~\cite{moveit} with the same start and end points, with the final trajectory chosen by the user~\cite{rosen2019}. This design has been shown to improve task completion time, precision, accuracy, usability, and task load compared to the baseline systems. Similarly, EVE uses an AR robot to communicate how a real robot might move with the collected waypoints. We also allow users to reselect the trajectory they desire by modifying waypoints. Unlike prior works, EVE is an iOS app, and the AR visualization is applied with a phone or a tablet. This form factor enables more people to collect demonstrations for training robots compared to wearable displays. Furthermore, instead of projecting the entire pose of the robot in AR, we move the AR robot directly to the goal pose due to the lack of safety concerns caused by physical collisions.

\section{EVE initial prototype}
\begin{figure}
    \centering
    \includegraphics[width=\linewidth]{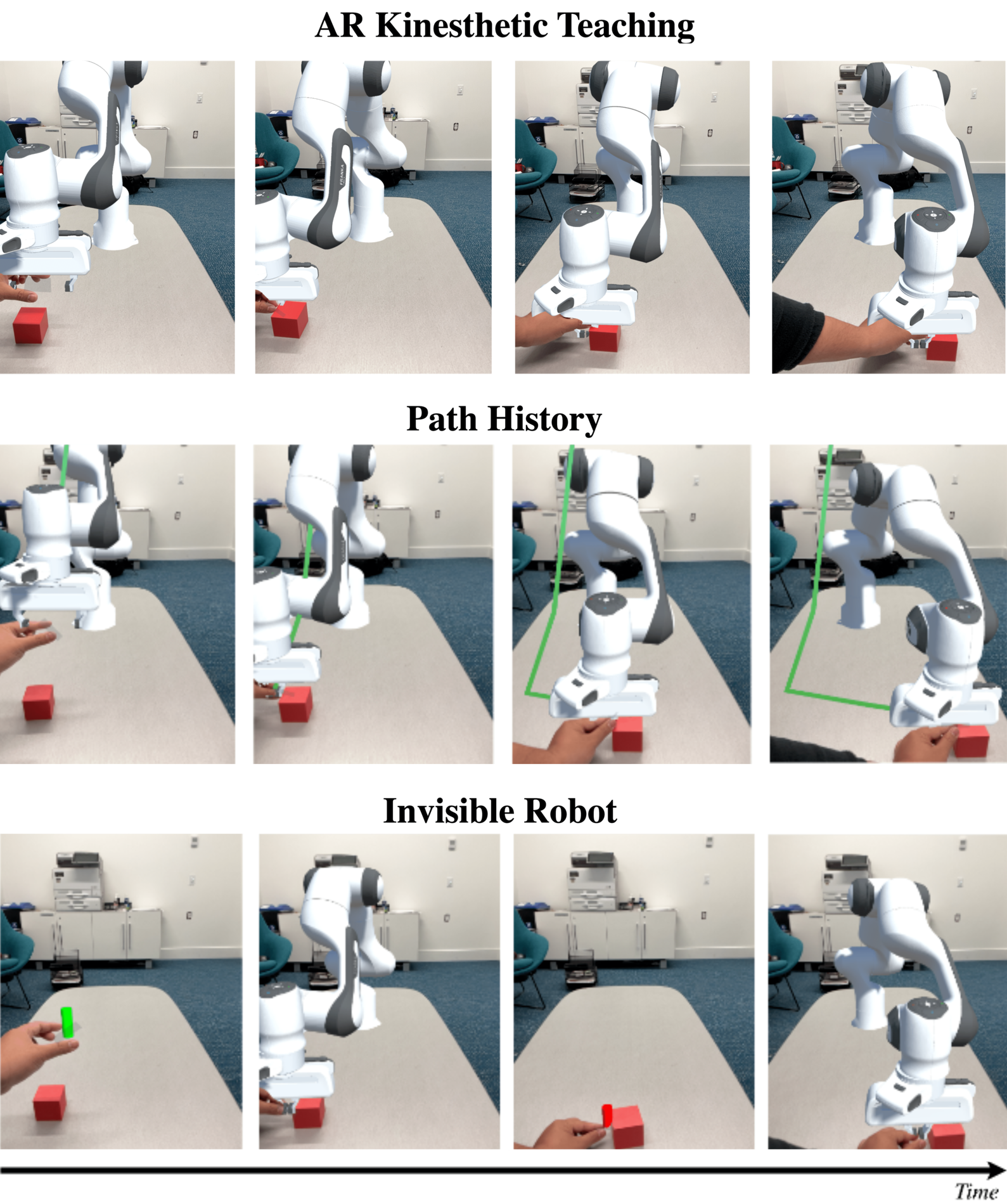}
    \caption{The initial prototype of EVE includes three new AR visualizations. \textit{AR kinesthetic teaching} allows the robot to track the user's hand movements in real-time. \textit{Path history} displays the trajectory along collected waypoints, enabling users to revert to previous waypoints. \textit{Invisible robot} allows users toggle off the robot's body, showing a cylinder that represents the robot's end effector position and gripper state.}
    \label{fig:EVE_formative}
    \Description{Three new AR visualizations in the formative study: AR Kinesthetic Teaching (top row), Path History (middle row), and Invisible Robot (bottom row). Each row consists of four images illustrating using the respective visualization to push the cube.}
\end{figure}

While prior work has compared common demonstration collection interfaces and AR visualizations, there is a lack of research investigating user preferences for AR-based demonstration collection systems. Informed by the usability issues of AR2-D2, we developed the initial EVE prototype with three new AR visualizations: \textit{AR kinesthetic teaching}, \textit{path history}, and \textit{invisible robot} (Figure \ref{fig:EVE_formative}).
Before we describe these visualizations, we first describe the user workflow for collecting trajectories.

\subsection{Trajectory collection in AR}
Our system enables users to collect AR trajectories for training real robots. After selecting a task for data collection, the user projects an AR robot into the real-world environment using the iOS app and guides the robot by specifying trajectories through hand movements. The application captures the trajectory of the AR robot by saving RGB-D data and waypoint coordinates. For instance, to train the robot to push a drawer, the user opens the app and directs the camera to the target location for the AR robot. By tapping the screen, they designate where the AR robot will appear. The user can then move the gripper by defining waypoints with their hands. The system captures and records RGB-D data along with waypoint coordinates during this interaction, saving the data to the device’s album for subsequent use in training the real robot policy.

While AR2-D2 demonstrated the possibility of using a similar workflow to collect robot trajectories, we identified several usability issues. Firstly, users must specify waypoints before pressing a button to move the AR robot. This can be unintuitive as users may struggle to visualize the robot's configuration at the waypoint. Besides, users cannot modify the collected waypoints, making it difficult to gather data for complex, long-horizon tasks without restarting the application. Lastly, when the AR robot is placed near the target object, it can occupy a significant portion of the device's screen, further complicating the data collection process. These usability challenges remained invisible in AR2-D2 because of their focus on simple, short-horizon tasks, such as picking up a plastic bowl and pressing a computer mouse.

\subsection{AR kinesthetic teaching (AR-KT)}
\textit{AR-KT} enhances system usability by synchronizing the robot's movements with the user's hand. To ensure natural robot motion, the system tracks the user’s hand position in real-time and leverages inverse kinematics to calculate the joint transformations necessary for robot movement. Prior work highlights that kinesthetic teaching generally surpasses teleoperation in system usability, collection time, and task success rate~\cite{Akgn2011RobotLF}, primarily due to the intuitiveness and immediate feedback of manually guiding the robot. Moreover, recent research~\cite{maric2024} has shown that human demonstration using a virtual marker excels in quality, time, and workload compared to kinesthetic teaching when drawing patterns. \textit{AR-KT} is similar to this approach as it allows users to move their hands freely without the need to manually adjust robot joint configurations to match desired positions.

\subsection{Path history}
\textit{Path history} leverages Unity’s Line Renderer component to display virtual trajectory lines between collected waypoint coordinates. Our inverse kinematics implementation uses the Jacobian matrix to relate joint velocities to end effector velocity and applies a damped least squares approach to compute joint angle updates. This iterative adjustment allows the end effector to move in near-straight-line paths toward the target pose (Figure~\ref{fig:EVE_formative}). Trajectory visualization is a well-established technique for enhancing motion intent communication, allowing users to preview a robot's planned path before execution. Prior research indicates that AR-based trajectory displays improve intent clarity and decrease task completion times~\cite{communicateAR,rosen2017,rosen2019}.

To facilitate real-time adjustments and flexibility, we implement a redo button that reverts the robot to its last saved waypoint and clears the associated trajectory. We choose this functionality over projecting the robot's goal pose before movement because our system is entirely AR-based, ensuring both safety and efficiency when moving the robot directly to the goal pose.

\subsection{Invisible robot}
The AR robot's 1:1 scale with the real Franka Panda robot often occludes the surroundings. To address this, \textit{invisible robot} renders the AR robot invisible and displays a cylinder to represent the position and orientation of the end effector. The cylinder's color changes based on the gripper's state: green when the gripper is open and red when the gripper is closed (Figure~\ref{fig:EVE_formative}). The color provides users with intuitive visual feedback about the gripper's current state without obstructing their view of the environment.

\section{Formative study}
\begin{table*}[t]
  \centering
  \includegraphics[width=1.0\linewidth]{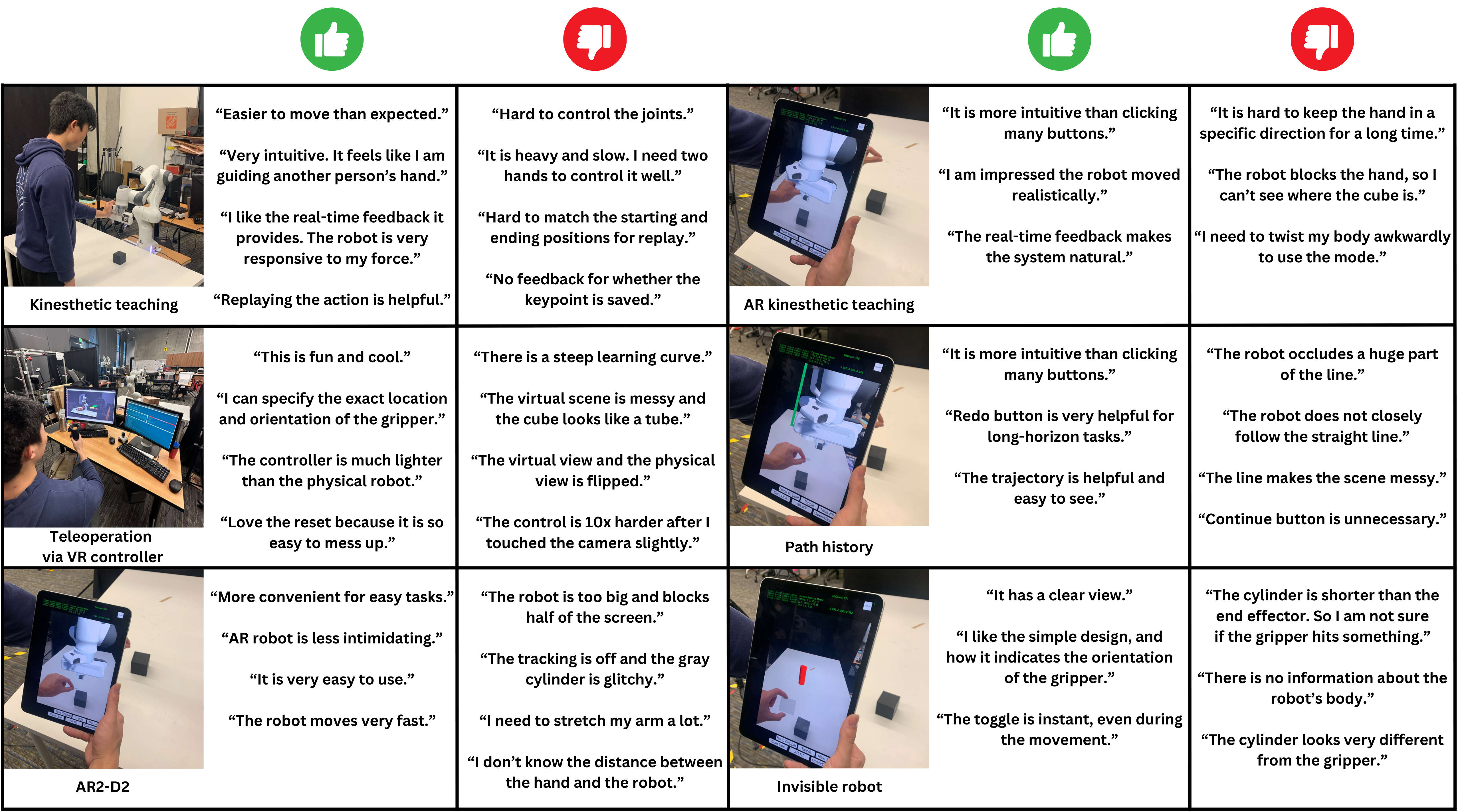}
  \caption{User comments about the six demonstration collection methods used in the formative study.}
  \label{fig:comments}
  \Description{User feedback on six demonstration collection interfaces in the formative study. The table includes an image of each interface alongside positive and negative comments from participants.}
\end{table*}

To study the utility of our initial prototype, we designed a formative study to 1) identify the challenges users encounter when collecting demonstrations using common collection interfaces and 2) evaluate the effectiveness of the new AR visualizations in aiding users with demonstration collection.

\subsection{Method}

\noindent\textit{\textbf{Participants.}} Since EVE aims to support everyday users, we recruited $10$ participants via snowball sampling. Participants were screened via a demographic questionnaire, which asked about their demographics and backgrounds, including age, gender, area of expertise, familiarity with robotics (5-point Likert scale), familiarity with demonstration collection interfaces (5-point Likert scale), and tasks they prefer the robot to help them with. We followed best practices from the HCI Guidelines for Gender Equity and Inclusivity~\cite{Scheuerman2020HCIGuidelines} to collect gender-related information. Participants ($\mu = 20.7$ years old, $\sigma = 1.7$ years), denoted as F01-F10, represented seven different areas of expertise and varying levels of familiarity in robotics and demonstration collection (Table~\ref{demographics}). The participants' familiarity with robotics ranged from unfamiliar to somewhat familiar: one person was unfamiliar, five were somewhat unfamiliar, three had neutral familiarity, and one was somewhat familiar. Regarding their familiarity with the demonstration collection, six participants were unfamiliar, and four were somewhat unfamiliar. Eight participants identified as male, while two participants identified as female.

\noindent\textit{\textbf{Procedure.}} The formative study took place on a university campus and lasted $50$ minutes. We obtained the Institutional Review Board's approval before the formative study, and the consent form was taken in person. All sessions were voice recorded for \textit{post hoc} analysis. Because we were interested in candid reactions, we did not tell participants that we created the additional AR visualizations. Each participant was compensated with food for the study.

The researchers began by demonstrating how to use each interface. Then, participants were given a few minutes to try each interface and ask any technical questions. Afterward, participants were given up to five minutes to try each interface by pushing, picking, and placing various cubes.

The formative study consisted of two parts. In part one, participants used kinesthetic teaching, teleoperation with an HTC Vive VR controller, and the AR2-D2 system. After the participants had tried data collection with the three state-of-the-art collection interfaces, we conducted a semi-structured interview about their experience, focusing on what features they liked and disliked about each interface. Although part one does not provide direct insight into an AR-based demonstration collection system, we included it to provide participants with a more comprehensive experience with robot data collection. 

In part two, participants performed the same task with \textit{AR-KT}, \textit{path history}, and \textit{invisible robot} visualizations. Aside from discussing the same questions in part one, we brainstormed additional features for the AR-based collection interface. Subsequently, participants completed a survey for System Usability Scale (SUS) scores and a form ranking the motion intent communication, user enjoyment, and preference for the AR2-D2 system and the new AR visualizations. Finally, we explained to the participant how the participant's data would be used to investigate the design for AR-based demonstration collection interfaces and answered any relevant questions.

\noindent\textbf{\textit{Analysis.}}
We analyzed two sources of data: interview transcripts and the post-task questionnaires. For the qualitative data, we used reflexive thematic coding to analyze the usability challenges faced by the participants with the current AR-based demonstration collection system. The first author, who facilitated all user study sessions, created an initial codebook by reviewing study transcripts. The entire team then collaboratively iterated on the codebook while checking for bias and coverage. With a final codebook consisting of $15$ codes, the team discussed the resulting themes. We selected and summarized several participants' quotes regarding what they liked and disliked about each collection interface (Figure~\ref{fig:comments}). For SUS scores, we converted survey responses, which are on a scale of 0-40 when summed, to a range between $0-100$\cite{brooke1996sus}\footnote{$(((Q1 + Q3 + Q5 + Q7 + Q9) - 5) + (25 - (Q2 + Q4 + Q6 + Q8 + Q10))) * 2.5$}. We calculated the mean and the standard deviation of the ranking forms. For all the scores, we conducted a Friedman test as an omnibus test with an appropriate number of Wilcoxon signed-rank tests corrected with Holm's sequential Bonferroni procedure for statistical significance. See Table \ref{table:formative_measurements} for a summary of quantitative results from the post-task questionnaires.

\begin{table}[h]
\centering
\fontsize{6}{6.8}\footnotesize
\renewcommand{\arraystretch}{2}
\begin{tabular}{m{2.3cm}>{\centering}m{1.11cm}>{\centering}m{1.11cm}>{\centering}m{1.11cm}>{\centering\arraybackslash}m{1.11cm}}
    \text{} & \textbf{AR2-D2} & \textbf{AR-KT} & \textbf{Path History} & \textbf{Invisible Robot} \\
    \toprule
    \textbf{Usability} (SUS) & $66.3\text{ }(20.3)$ & $67.3\text{ }(12.3)$ & $\textcolor{darkgreen}{\textbf{70.0 (13.0)}}$ & $68.3\text{ }(15.7)$ \\
    \textbf{Motion Intent} (rank) & $2.9\text{ }(1.2)$ & $2.4\text{ }(1.0)$ & $\textcolor{darkgreen}{\textbf{1.9 (0.9)}}$ & $2.8\text{ }(1.3)$ \\
    \textbf{Enjoyment} (rank) & $3.5\text{ }(1.0)$\textasciicircum & $2.4\text{ }(1.0)$ & $\textcolor{darkgreen}{\textbf{2.0 (0.7)}}$\textasciicircum & $2.1\text{ }(1.3)$ \\
    \textbf{Preference} (rank) & $3.5\text{ }(1.0)$\textasciicircum & $2.7\text{ }(0.8)$ & $\textcolor{darkgreen}{\textbf{1.5 (0.7)}}$\textasciicircum & $2.3\text{ }(1.1)$ \\
    \hline
    \text{} & \text{} & \text{} & \text{} & \text{}
\end{tabular}
\caption{\textit{Path history} outperformed other AR-based collection methods in usability, motion intent communication, user enjoyment, and user preference. The mean and standard deviation of all measurements are shown. Usability is rated on a scale from 0 to 100, with higher scores indicating better usability. Motion intent communication, user enjoyment, and user preference are ranked from 1 to 4, with lower ranks indicating better performance. Statistical significance is indicated by one caret (\textasciicircum) for $p < 0.05$.}
\label{table:formative_measurements}
\Description{Mean and standard deviation of usability, motion intent, enjoyment, and preference of AR2-D2, AR Kinesthetic Teaching, Path History, and Invisible Robot.}
\end{table}

\subsection{Findings}
The \textit{path history} visualization significantly outperformed the baseline AR2-D2 system in user enjoyment ($p=0.0059$) and user preference ($p=0.0059$). Notably, the \textit{path history} visualization achieved the highest median scores and rankings across all measured variables: SUS ($\mu = 70.0$, $\sigma = 13.0$), motion intent communication ($\mu = 1.9$, $\sigma = 0.9$), user enjoyment ($\mu = 2.0$, $\sigma = 0.7$), and preference ($\mu = 1.5$, $\sigma = 0.7$). A Friedman test found no statistically significant differences between the four AR visualizations in the SUS scores ($\chi^2(3) = 2.6$, $p=0.5$) and rankings of motion intent communication ($\chi^2(3) = 3.7$, $p=0.3$). These findings suggest that participants found the \textit{path history} visualization the most usable, enjoyable, and preferable among the four tested AR visualizations, even though it did not significantly outperform the others in all aspects.

Interestingly, these results contradicted our hypotheses. Given the benefits of direct manipulation and real-time visual feedback, we had expected the \textit{AR-KT} visualization to achieve the highest scores on system usability and motion intent communication compared to other visualizations. The contradiction may be due to the AR robot's inability to accurately track the hand when it moves quickly due to the computational limit of the device and the inaccuracy of the device's LiDAR camera. We also anticipated that the \textit{invisible robot} visualization would receive the lowest scores on system usability and motion intent communication due to the lack of joint information. The contradiction could be attributed to the approximate straight-line motion of the AR robot, which allows users to easily predict the movement of the joints.

In addition, we distilled the challenges users faced with the current AR visualizations into the following six based on the semi-structured interviews:
\renewcommand{\labelenumi}{\textbf{C\arabic{enumi}}}
\begin{enumerate}
    \item \textbf{Ambiguous spatial position of the hand and the robot:} Four participants (F03, F04, F08, F10) expressed difficulty interpreting the 3D environment via a 2D iPad screen. The lack of depth perception made it challenging for users to accurately gauge the relative positions of their hands and the AR robot. This ambiguity led to inaccuracies in the collected demonstrations. F03 also complained about the hand passing through the robot, resulting in an unnatural collection.
    \item \textbf{Lack of knowledge of joint constraints:} Five participants (F04, F05, F06, F08, F10) reported a lack of information about the joint limits. This problem, coupled with the lack of depth perception via the iPad screen, causes the robot to move unexpectedly when the specified point with the hand is out of the joint limit.
    \item \textbf{Imprecise trajectory visualization:} The \textit{path history} visualization displays a straight green line between the robot's end effector coordinate and the user's hand coordinate. Although three participants (F05, F06, F10) stated that the trajectory visualization aids their understanding of the robot's motion intent, three other participants (F02, F04, F07) have proposed to include a more accurate trajectory.
    \item \textbf{Absence of feedback regarding the collection efficacy:} Five participants (F01, F04, F08, F09, F10) complained about the lack of collision detection, which prevented them from verifying whether the collected trajectory was feasible. F01 claimed that while AR is ``useful and fun," it does not make sense for the AR robot to pass through all the obstacles.
    \item \textbf{Obstructive robot design:} The AR robot's 1:1 scale correspondence with the physical Franka Panda robot frequently resulted in the occlusion of relevant objects within the scene. Four participants (F01, F05, F07, F09) reported that the AR robot's size significantly impeded the data collection. Specifically, F01 expressed frustration regarding the robot's obstruction of the line, while F09 noted that the robot's presence obscured a substantial portion of the iPad's display. 
    \item \textbf{Inconsistent hand tracking:} Inconsistencies in hand tracking emerged as a significant concern among participants. Five individuals (F01, F02, F03, F06, F08) highlighted the presence of inaccuracies and latency in the hand tracking system employed in the AR environment. Furthermore, F02 noted that the gray line renderer, utilized in AR2-D2 to visualize the tracked coordinates of the user's fingers, is visually distracting.
\end{enumerate}

\section{EVE final prototype}
\begin{figure*}
    \centering
    \includegraphics[width=\linewidth]{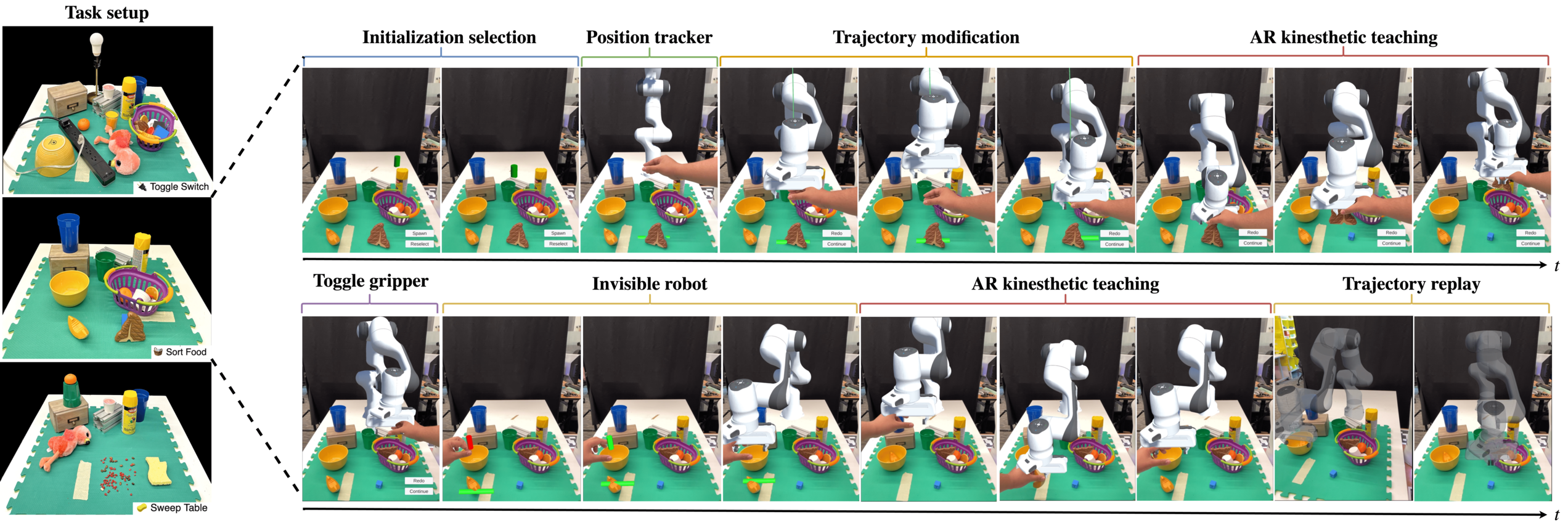}
    \caption{Overview of the evaluation user study and prototype 2 of EVE. \emph{Left:} The study included three common tabletop tasks: toggling a switch, sorting food, and sweeping the table. \emph{Right:} An example of data collection for the food sorting task using EVE final prototype.}
    \label{fig:EVE_evaluation}
    \Description{The setup of the three tabletop tasks used in the evaluation study (left). An example collection of the sort food task using EVE prototype 2 (right).}
\end{figure*}

Informed by formative study findings and our own experiences using EVE, we upgraded EVE's prototype with seven additional system changes to address the usability challenges. We describe these advancements below:

\subsection{Hand position projection}
The 2D display of the smartphone or tablet makes position perception of the hand and the robot in 3D space challenging \textbf{(C1)}. Prior work addresses this problem with a virtual cast shadow of a flying drone~\cite{PinpointFly}. We imitated this approach by representing the shadow of the AR robot's end effector with a green line on the surface of the table. To further enhance motion intent communication, the length and width of the projected line were adjusted to match the dimensions of the robot's end effector. This feature also aids users become familiar with the robot’s morphology and estimate what positions would be unreachable \textbf{(C2)}.

\subsection{Joint constraints signifier}
To address user concerns about joint constraints \textbf{(C2)}, we adhered to the joint space limits specified in the Franka Emika Panda robot documentation~\cite{panda_param}. This allowed us to define a range of reachable coordinates based on the AR robot's initial position. When the user's hand is outside this reachable range, all buttons turn red (otherwise white) and are deactivated, preventing any changes to the robot's state.

\subsection{Realistic trajectory}
Three participants (F02, F04, F07) indicated that the straight-line trajectory visualization did not effectively communicate the robot's motion intent \textbf{(C3)}. In response, the system was modified to display the precise, curved trajectory of the end effector. Moreover, \textit{path history} is now the default visualization for EVE, as it demonstrated superior performance compared to other AR visualizations in the formative user study. We disabled trajectory visualization during \textit{AR-KT} because the user is controlling the robot in real-time.

\subsection{Trajectory replay}
The lack of feedback regarding collision detection and path feasibility significantly hinders the system's usability \textbf{(C4)}. Detecting collisions of the AR robot with the physical objects or planning a trajectory for the AR robot to avoid physical obstacles typically requires constructing a depth map of the environment, which can be computationally expensive for the iPad. After brainstorming potential solutions with the formative study participants, we added a replay feature that demonstrates the collected trajectory and changes in the gripper state using a partially transparent robot. This feature makes it easier to determine whether the robot collided with a physical object during collection. Additionally, this implementation was inspired by some participants' preferences (F02, F03) for the replay feature available in kinesthetic teaching. The system linearly interpolates 20 waypoints for the trajectory collected with \textit{AR-KT}, allowing users to modify waypoints and replay the trajectory.

\subsection{Dynamic camera view}
To address the issue of the AR robot occupying a significant portion of the screen \textbf{(C5)}, users need to view the robot from different angles. Prior work enhanced situational awareness using an interactive third-person perspective from a second, spatially coupled drone~\cite{3Piloting}. After discussing with a formative study participant (F05), we decided to allow users to move the device to inspect the 3D world, as setting up additional smartphones or tablets is costly and inconvenient. However, collecting demonstrations from multiple camera angles is not ideal because it is crucial to maintain hand-eye calibration between the robot's actions and the 3D voxels for successful real-world deployment. Different camera angles might provide conflicting or misleading information about distances and spatial relationships, which can disorient the robot's understanding and actions. To address this, we save the initial camera extrinsic values after the robot is instantiated and make the buttons disappear when the current extrinsic values deviate from the saved values by more than a small threshold. This approach enables users to freely inspect the scene from multiple angles while ensuring the demonstration is collected from a consistent camera angle. This feature also helps users become familiar with the robot's morphology and estimate which positions might be unreachable \textbf{(C2)}.

\subsection{Accurate gripper control}
In AR2-D2, a gray line is rendered between the thumb and index fingers, allowing users to toggle the gripper opening by adjusting the distance between their fingers. However, due to limitations in the iPad camera and ARKit's HandTrackingProvider, hand tracking is inconsistent, making the toggling of the gripper state unpredictable \textbf{(C6)}, as reported by five participants from the formative study (F01, F02, F03, F06, F08). To maintain the system's ubiquity, we opted not to modify the hardware. Instead, we resolved the issue of the flashing gray line renderer by removing it and incorporating a button to accurately toggle the gripper opening. This adjustment ensures a more reliable and predictable user experience when controlling the gripper state.

\subsection{Robot instantiation reselection}
Our experience with the system, along with feedback from one participant (F09), suggests that the inability to change the robot's instantiation point hinders usability, as it requires restarting the application if the robot is placed in an undesired location. To address this issue, we revised the robot instantiation process into two steps. First, a green cylinder appears on the screen to indicate where the robot will spawn. The user is then given two choices: instantiate the robot at the position indicated by the cylinder or reselect a position for the green cylinder. This ability to reselect the instantiation point for the AR robot significantly enhances the system's usability.

\section{Evaluation}
We conducted a user study to evaluate the effectiveness of EVE compared to baseline interfaces for three common tabletop tasks. These tasks were designed based on the survey from BEHAVIOR-1K~\cite{Li2024BEHAVIOR1KAH} asking participants, ``What do you want robots to do for you?" We included obstacles in all tasks to simulate real-world environments where tables are often cluttered. The three tasks were as follows:
\begin{itemize}
    \item \textbf{Toggle switch:} Participants guided the robot to push a switch, turning on a light bulb. This task is more challenging than the button press task from the AR2- D2 paper, as it requires millimeter-level precision for the robot’s end-effector to touch a specific point on the switch.
    \item \textbf{Sort food:} Participants controlled the robot to grab a melon and place it inside a yellow bowl, then pick up a steak and put it into a basket. This task simulated sorting different items into specific locations.
    \item \textbf{Sweep table:} Participants directed the robot to sweep beans to a location marked by tape with a sponge. This task simulated cleaning debris from a table.
\end{itemize}
The setups for these tasks are illustrated in Figure~\ref{fig:EVE_evaluation}.

\subsection{Method}

\noindent\textbf{\textit{Participants.}}
To ensure a diverse area of expertise, we recruited $14$ participants via snowball sampling. Participants were screened using the same demographic questionnaire from the formative study. To reduce bias, we ensured that none of the participants participated in the formative study. We followed best practices from the HCI Guidelines for Gender Equity and Inclusivity~\cite{Scheuerman2020HCIGuidelines} to collect gender-related information. Participants ($\mu = 22.6$ years old, $\sigma = 2.6$ years), denoted as E01-E14, represented 11 different areas of expertise and varying levels of familiarity in robotics and demonstration collection (Table~\ref{demographics}). The participants' familiarity with robotics ranged from unfamiliar to familiar: three people were unfamiliar, six people were somewhat unfamiliar, three people had neutral familiarity, and two people were familiar. Regarding their familiarity with the demonstration collection, eight people were unfamiliar, three people were somewhat unfamiliar, two people had neutral familiarity, and one person was somewhat familiar. $11$ participants identified as male, while three participants identified as female.

\noindent\textbf{\textit{Procedure.}}
The laboratory study was conducted on a university campus, with each session lasting $40$ minutes. Consent forms were collected in person. To ensure candid reactions, participants were not informed that we developed EVE. Participants were compensated with food for their participation in the study.

Each session began with a brief tutorial on all the collection interfaces, including kinesthetic teaching, teleoperation with an HTC Vive VR controller, AR2-D2, and EVE. Participants were informed about the goals of the tasks but were not shown how to accomplish them to avoid bias. After a $30$-second practice session for each interface, participants began the task collection.

A collection was deemed successful if the participant completed the task without any collisions within three minutes, excluding setup times. If a collision occurred and the three-minute limit had not been reached, participants were allowed to retry. For kinesthetic teaching and teleoperation, a collision was defined as the robot hitting objects that were not the target objects. For AR2-D2 and EVE, collisions were manually assessed by the study proctor by checking if any part of the AR robot intersected with physical objects during collection and task replay. This assessment strategy was chosen as the alternative would require expensive hardware setup with multiple cameras and constructing a depth map is computationally expensive for iPad.

Optimal data collection for AR2-D2 and EVE required that the distance between the AR robot and the scene match that of the physical robot and its surroundings. This is a common assumption in robotics research due to the fixed camera location relative to the robot. To ensure data accuracy, we marked a point on the table and positioned the AR robot at the same spot before each collection. future work could utilize physical reference markers to calibrate the camera and the collected data, thereby removing this assumption.

Success rates and remaining time for each task were recorded for all interfaces. The remaining time was used instead of total time to maintain consistency with the representation of success rates, where higher values indicate better performance. Data collection time for each interface was limited to three minutes, as it is challenging to collect a successful demonstration using a virtual reality controller for teleoperation.

Upon completing the task collection with all interfaces, participants filled out the SUS survey and a form ranking motion intent communication, user enjoyment, and overall preference for the interfaces. We aimed to measure $10$ collection attempts for each task. To prevent fatigue from using four interfaces per task, participants could stop at any time, resulting in each participant collecting between one and three tasks. All participants completed a NASA TLX survey post-study to assess the workload. Finally, we explained to the participant how the participant's data would be used to investigate the design for AR-based demonstration collection interfaces and answered any relevant questions.

\noindent\textbf{\textit{Analysis.}}
We analyzed the results from the SUS survey and ranking forms with the same method used for the formative study result analysis. See Table~\ref{table:evaluation_measurements} for a summary of the quantitative results from the post-task questionnaires. Additionally, we calculated the overall task success rate and determined the mean and standard deviation for the remaining time and the NASA TLX survey.

\subsection{Findings}
EVE achieved the highest task success rate and performed comparably to kinesthetic teaching in terms of task completion time, system usability, motion intent communication, user enjoyment, and overall user preference ($mean_{p}=0.30$). As expected, Kinesthetic teaching attained the highest SUS score and was ranked highest in motion intent, user enjoyment, and overall preference. However, there was no statistically significant difference between EVE and kinesthetic teaching, suggesting that EVE may be a viable alternative for collecting trajectory data. This could be due to the ability to clearly inspect the environment during collection and the intuitive control of guiding the robot with the hand (\textit{AR-KT}). Our results support previous findings \cite{wrede2013user, Akgn2011RobotLF}, which showed that kinesthetic teaching outperforms VR controllers in accuracy, speed, and user preference. EVE outperformed AR2-D2 in all measurements. The NASA TLX survey post-study indicated a low to medium workload for all users after the collection ($\mu = 2.57$, $\sigma = 2.45$), further reinforcing the validity of our findings.

\begin{figure}[t]
  \centering
  \includegraphics[width=1.0\linewidth]{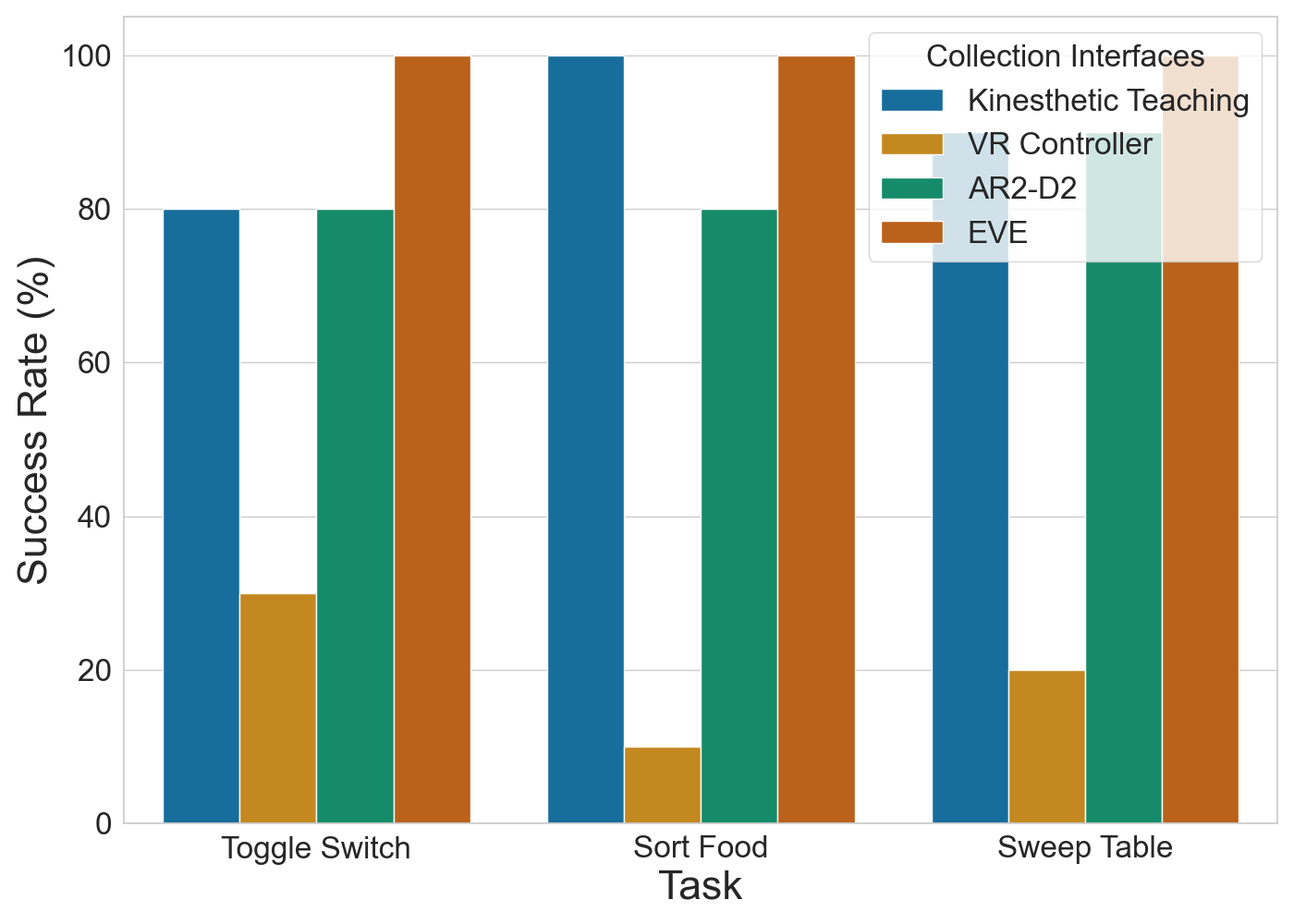}
  \caption{The mean task success rate (\%) for all interfaces in the evaluation user study, which included three tasks with 10 trials each, indicates that EVE achieved the highest success rate across all tasks.}
  \label{fig:success_rate}
  \Description{Bar graphs titled “Task Success Percentage” showing the three tabletop tasks in the evaluation study on the X axis and the mean success rate (\%) of the four collection interfaces on the Y axis.}
\end{figure}

\begin{figure}[t]
  \centering
  \includegraphics[width=1.0\linewidth]{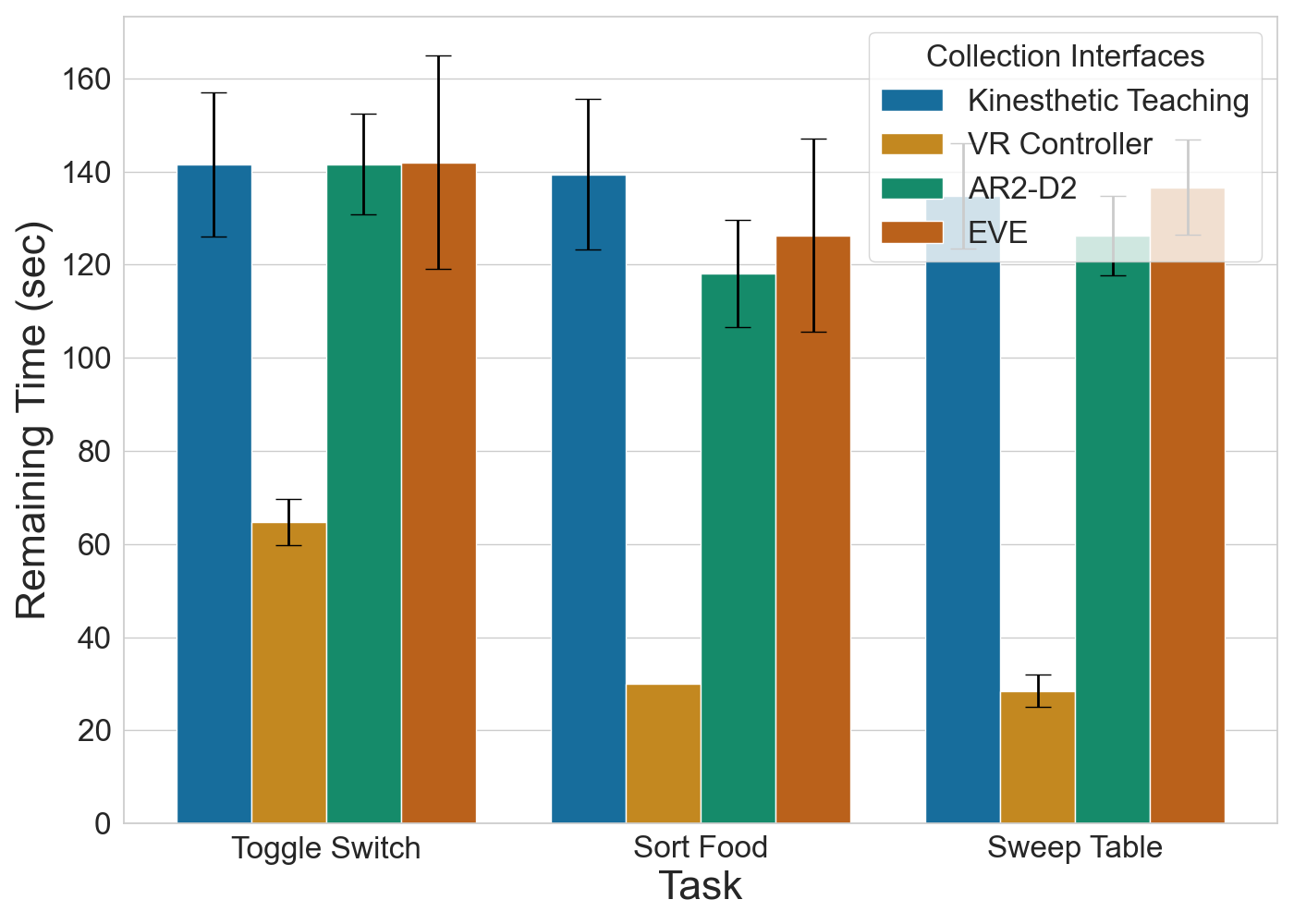}
  \caption{The mean and the standard deviation of the remaining time (seconds) for successfully completing one demonstration for each task. EVE performed comparably to kinesthetic teaching, with an average difference of 5.1 seconds across the three tasks.}
  \label{fig:remaining_time}
  \Description{Bar graphs titled “Task Remaining Time” showing the three tabletop tasks in the evaluation study on the X axis and the mean and standard deviation of the remaining time (sec) of the four collection interfaces on the Y axis. The standard deviation of the Teleoperation via VR controller for the sort food task is missing because only one person succeeded in the task collection.}
\end{figure}

\begin{table}[ht]
\centering
\fontsize{6}{6.8}\footnotesize
\renewcommand{\arraystretch}{2}
\begin{tabular}{b{2.17cm}>{\centering}m{1.14cm}>{\centering}m{1.14cm}>{\centering}m{1.14cm}>{\centering\arraybackslash}m{1.14cm}}
    {\large \textbf{Overall}} & \textbf{{\scriptsize Kinesthetic} Teaching} & \textbf{Teleoperation} & \textbf{AR2-D2} & \textbf{EVE} \\
    \toprule
    \textbf{Usability} (SUS) & $\textcolor{darkgreen}{\textbf{86.0 (17.0)}}$ & $\text{27.3 (17.8)}$* & $\text{59.0 (20.5)}$* & $\text{77.8 (17.0)}$ \\
    \textbf{Motion Intent} (rank) & $\textcolor{darkgreen}{\textbf{1.4 (0.7)}}$ & $\text{3.3 (0.8)}$* & $\text{3.4 (0.7)}$* & $\text{1.9 (0.8)}$ \\
    \textbf{Enjoyment} (rank) & $\textcolor{darkgreen}{\textbf{1.5 (0.8)}}$ & $\text{3.2 (1.0)}$* & $\text{3.3 (0.7)}$* & $\text{2.0 (0.8)}$ \\
    \textbf{Preference} (rank) & $\textcolor{darkgreen}{\textbf{1.7 (0.9)}}$ & $\text{3.4 (0.9)}$* & $\text{3.1 (0.7)}$* & $\text{1.8 (0.7)}$ \\
    \hline
    \text{} & \text{} & \text{} & \text{} & \text{} \\
    
    {\large \textbf{Toggle Switch}} & \textbf{{\scriptsize Kinesthetic} Teaching} & \textbf{Teleoperation} & \textbf{AR2-D2} & \textbf{EVE} \\
    \toprule
    \textbf{Usability} (SUS) & $\textcolor{darkgreen}{\textbf{83.8 (18.0)}}$ & $\text{31.3 (21.5)}$* & $\text{66.0 (20.2)}$* & $\text{82.8 (13.9)}$ \\
    \textbf{Motion Intent} (rank) & $\textcolor{darkgreen}{\textbf{1.4 (0.7)}}$ & $\text{3.4 (1.0)}$ & $\text{3.2 (0.6)}$* & $\text{2.0 (0.8)}$ \\
    \textbf{Enjoyment} (rank) & $\text{2.0 (1.1)}$ & $\text{3.0 (1.2)}$ & $\text{3.1 (0.9)}$* & $\textcolor{darkgreen}{\textbf{1.9 (1.0)}}$ \\
    \textbf{Preference} (rank) & $\text{2.2 (1.0)}$ & $\text{3.3 (1.1)}$ & $\text{2.9 (0.9)}$* & $\textcolor{darkgreen}{\textbf{1.6 (0.8)}}$ \\
    \hline
    \text{} & \text{} & \text{} & \text{} & \text{} \\
    
    {\large \textbf{Sort Food}} & \textbf{{\scriptsize Kinesthetic} Teaching} & \textbf{Teleoperation} & \textbf{AR2-D2} & \textbf{EVE} \\
    \toprule
    \textbf{Usability} (SUS) & $\textcolor{darkgreen}{\textbf{88.0 (17.1)}}$* & $\text{26.5 (15.5)}$* & $\text{54.0 (20.6)}$* & $\text{71.0 (17.5)}$ \\
    \textbf{Motion Intent} (rank) & $\textcolor{darkgreen}{\textbf{1.3 (0.5)}}$ & $\text{3.1 (0.7)}$ & $\text{3.7 (0.5)}$* & $\text{1.9 (0.7)}$ \\
    \textbf{Enjoyment} (rank) & $\textcolor{darkgreen}{\textbf{1.2 (0.4)}}$* & $\text{3.1 (1.1)}$ & $\text{3.5 (0.5)}$* & $\text{2.2 (0.6)}$ \\
    \textbf{Preference} (rank) & $\textcolor{darkgreen}{\textbf{1.2 (0.4)}}$ & $\text{3.5 (0.7)}$* & $\text{3.4 (0.5)}$* & $\text{1.9 (0.6)}$ \\
    \hline
    \text{} & \text{} & \text{} & \text{} & \text{} \\
    
    {\large \textbf{Sweep Table}} & \textbf{{\scriptsize Kinesthetic} Teaching} & \textbf{Teleoperation} & \textbf{AR2-D2} & \textbf{EVE} \\
    \toprule
    \textbf{Usability} (SUS) & $\textcolor{darkgreen}{\textbf{86.3 (17.5)}}$ & $\text{24.3 (17.1)}$* & $\text{56.8 (20.9)}$* & $\text{80.0 (18.7)}$ \\
    \textbf{Motion Intent} (rank) & $\textcolor{darkgreen}{\textbf{1.6 (0.8)}}$ & $\text{3.3 (0.8)}$ & $\text{3.3 (0.8)}$* & $\text{1.8 (0.8)}$ \\
    \textbf{Enjoyment} (rank) & $\textcolor{darkgreen}{\textbf{1.4 (0.7)}}$ & $\text{3.5 (0.8)}$* & $\text{3.2 (0.6)}$* & $\text{1.9 (0.7)}$ \\
    \textbf{Preference} (rank) & $\textcolor{darkgreen}{\textbf{1.6 (0.8)}}$ & $\text{3.5 (0.8)}$* & $\text{3.1 (0.7)}$* & $\text{1.8 (0.8)}$ \\
    \hline
    \text{} & \text{} & \text{} & \text{} & \text{} \\
\end{tabular}
\caption{EVE performed comparably to kinesthetic teaching in usability, motion intent communication, user enjoyment, and user preference. The mean and standard deviation of all measurements are shown. Usability is rated on a scale from 0 to 100, with higher scores indicating better usability. Motion intent communication, user enjoyment, and user preference are ranked from 1 to 4, with lower ranks indicating better performance. One asterisk (*) indicates statistical significance ($p < 0.05$) when compared to the EVE system.}
\label{table:evaluation_measurements}
\Description{Mean and standard deviation of usability, motion intent, enjoyment, and preference of Kinesthetic Teaching, Teleoperation via VR controller, AR2-D2, and EVE.}
\end{table}

Below are observations and feedback on using different collection interfaces for the tasks:

\noindent\textit{\textbf{Toggle switch.}}
This task involves navigating the AR robot's end effector around physical obstacles and accurately pressing the switch. With kinesthetic teaching, users appreciated the method's simplicity and the precise control of the end effector. However, some users needed to rotate the robot's body during data collection to avoid collisions, which increased manual effort. Teleoperation with a VR controller proved challenging, with most users failing to collect successful demonstrations due to the unintuitive controls and the difficulty in accurately perceiving the switch's position from the point cloud representation of the scene displayed on a 2D monitor. With AR2-D2, users frequently failed due to difficulties in interpreting the 3D position of the AR robot, resulting in demonstrations where the end effector touched points near the switch rather than the switch itself. Conversely, all users successfully collected demonstrations with EVE. Most users utilized the \textit{AR-KT} feature, aided by hand position projection, to guide the AR robot to the switch.

\noindent\textit{\textbf{Sort food.}}
This is a long-horizon task requiring users to control a robot to pick up and place multiple objects. With kinesthetic teaching, many users appreciated the precise control of the gripper for pick-and-place actions. In contrast, teleoperation posed challenges, as most users struggled to pick up food due to positional discrepancies between the point cloud representation and the actual scene. With AR2-D2, users found it difficult to control the gripper for grasping and releasing objects. Additionally, many users failed to collect successful demonstrations because they assumed the AR robot would automatically avoid physical obstacles to reach the specified waypoint. Conversely, all users successfully collected demonstrations with EVE. Most users rotated the iPad camera to prevent collisions between the AR robot's body and the blue cup.

\noindent\textit{\textbf{Sweep table.}}
This task required users to control a robot to grab a sponge and perform a series of sweeping actions. Many users preferred kinesthetic teaching for its simple control, allowing them to directly manipulate the robot’s end effector and receive haptic feedback when the end effector contacted the table surface. With the teleoperation system, users faced challenges due to latency and the precision needed to control the end effector with a VR controller. They struggled to maintain a consistent direction of movement, often resulting in collisions between the end effector and the table. With AR2-D2, users found it difficult to move the robot’s end effector horizontally, as specifying waypoints with the same vertical distance from the table surface was challenging. With EVE, most users favored using \textit{AR-KT} to specify waypoints, as they found it easier to maintain the hand's height during a continuous movement.

\subsection{Real world deployment}
To assess the effectiveness of the collected demonstrations, we trained robot policies for the \textit{toggle switch} task using data gathered by AR2-D2 and EVE.

\noindent\textit{\textbf{Training procedure.}}
Like AR2-D2, EVE collects RGB-D data and waypoint coordinates during the data collection process. After collecting the data, we used Perceiver-Actor (PerAct)~\cite{peract} to train a transformer-based language-guided behavior cloning policy. PerAct takes a 3D voxel observation and a language goal \textit{(v, l)} as inputs and produces discretized outputs for the translation, rotation, and gripper state of the end-effector. These outputs, in conjunction with a motion planner, enable the execution of tasks specified by the language goal. We trained the policy using $6$ demonstrations over $30,000$ iterations for each interface.

\noindent\textit{\textbf{Results.}}
The trained policies were evaluated with $30$ task rollouts for the \textit{toggle switch} task. The policy trained with EVE-collected data achieved a $30\%$ success rate, compared to a $20\%$ success rate with AR2-D2-collected data. We emphasize that the \textit{toggle switch} task is more challenging than the button press task from the AR2-D2 paper, as it requires millimeter-level precision for the robot’s end-effector to touch a specific point on the switch to turn on a light bulb. The higher success rate with EVE-collected data may be attributed to the improved depth perception during collection, allowing users to adjust waypoints for more accurate trajectories. We have included a video demonstrating a successful rollout of the \textit{toggle switch} task and the 3D voxel observations generated with data collected using AR2-D2 and EVE on our website\footnote{\href{https://junwang0510.github.io/EVE/}{https://junwang0510.github.io/EVE/}}.

\section{Discussion}
In this section, we reflect on insights from the development and evaluation of EVE. We also discuss future opportunities for research exploring AR-based demonstration collection interfaces for robotics.

\subsection{Implications}
\noindent\textbf{\textit{EVE enables personalized robot training in everyday environments.}}
EVE empowers users to gather demonstrations in diverse settings beyond traditional laboratories, exposing robots to a variety of objects and environments tailored to their specific needs. As robot hardware becomes more affordable and household robots more common, EVE puts the power of robot training directly into the hands of users, allowing them to collect personalized demonstrations for specific household tasks.

\noindent\textbf{\textit{EVE is compatible with various robot platforms and motion planners.}}
EVE enables users to quickly adapt the system to different robot platforms. Since each program component is modular, users can gather data for another type of robot by replacing the robot prefab in Unity with another model, such as the UR5 robot. Additionally, since EVE collects waypoint data in a standard format compatible with common motion planners, it facilitates the training of manipulation policies for different robots without needing to modify the collected data. This adaptability extends EVE's applicability to a wide range of robotic systems, making robot training more accessible to a broader audience.

\subsection{Limitations and future work}
\noindent\textbf{\textit{Bimanual robotic manipulation.}}
Many real-world tasks, such as washing dishes, ironing clothes, and assembling furniture, require the use of both hands. However, EVE currently does not support bimanual manipulations, as users need to press buttons during the collection process and it is uncomfortable to collect demonstrations with an iPad positioned in front of the body. Future research could address this limitation by exploring the use of AR head-mounted displays. This approach would allow users to control both robot arms with their hands while using voice commands to trigger specific operations. Additionally, these technologies enable the implementation of better depth cues, such as realistic shadows and dynamic lighting, to improve the user's ability to interact with the AR environment accurately. However, implementing bimanual controls may present additional challenges, such as accurately tracking both hands during occlusion.

\noindent\textbf{\textit{Collection for mobile navigation robots.}}
The real world also involves tasks that require mobile robots, such as cleaning the floor and watering plants. Extending EVE to support demonstration collection for mobile robots is a promising direction for future research. This can be achieved by integrating a 3D model of the mobile robot into Unity and developing intuitive gesture controls for its movement. For instance, users could draw the desired trajectory for the robot on an iPad, similar to the TrajectoryFly feature in PinpointFly~\cite{PinpointFly}. The system could also utilize the device's camera and inertial measurement unit (IMU) sensors to create a detailed map of the environment while tracking the robot's pose, thereby enabling effective data collection for mobile robots.

\noindent\textbf{\textit{Multi-view data collection.}}
EVE is limited to single-viewpoint data collection to maintain accurate hand-eye calibration between the robot's actions and the corresponding 3D voxel representation, which is essential for successful real-world deployment. However, multi-view data collection could be helpful in scenarios where objects are occluded from one perspective but visible from another, or when the demonstrator needs to maintain a more comfortable posture by collecting trajectories from different angles. Future work could explore the use of SLAM techniques~\cite{durrant2006slam} to track the iPad LiDAR's pose and build a 3D map of the workspace, allowing the demonstrator to collect trajectories relative to this 3D map.

\noindent\noindent\textbf{\textit{Dynamic tasks.}}
To enhance EVE's capability to perform tasks in unstructured environments, we integrated motion planning algorithms such as Fast-RRT and RRT* by leveraging LiDAR-generated octree maps. However, the system failed to operate in real-time due to the iPad's limited computational power. Future research could experiment with various edge devices and develop more efficient motion planners.

\noindent\textbf{\textit{Hardware.}}
We acknowledge that the reliance on specific hardware (iOS devices with AR capabilities) might limit the user base. Future work could evaluate EVE on different devices (e.g., Android devices, 3D glasses) or in less ideal task environments to test the generalizability of our design and findings. Additionally, the iPad's LiDAR camera occasionally results in inaccuracies in the end effector's movement and insufficient anchoring of the robot base on the surface. To address this issue and improve tracking accuracy and robustness, future work could implement sensor fusion techniques by combining data from multiple sensors, such as the iPad's camera, IMU, and other cameras.

\section{Conclusion}
We present EVE, an iOS application that democratizes robot data collection through human-centric AR visualizations. EVE empowers users to collect, correct, and inspect trajectory data to train robots for personalized tasks. In our evaluation user study, EVE outperformed three state-of-the-art interfaces in terms of success rate and matched the performance of kinesthetic teaching in completion time, usability, motion intent communication, enjoyment, and user preference. We hope EVE will catalyze future work in AR-based demonstration collection for robotics and expedite the incorporation of robots into daily lives.

\begin{acks}
We are grateful to Abhimanyu Saighal for his assistance in preprocessing the collected demonstrations for training robot policies. We also thank Sebastin Santy and Zixian Ma for their valuable feedback on the paper.
\end{acks}

\bibliographystyle{ACM-Reference-Format}
\bibliography{references}

\appendix
\clearpage
\onecolumn

\section{Study Participants Demographics} \label{demographics}
\begin{table*}[h]
\centering
\begin{tabular}{lllp{3.1cm}lp{3cm}l}
\toprule
\textbf{\#} & \textbf{\textit{f}} & \textbf{\textit{e}} & \textbf{Area of Expertise} & \textbf{Familiarity (Robotics)} & \textbf{Familiarity (Demonstration Collection)} & \textbf{Tasks}\\
\midrule
1 & {F01} & - & Marketing & Somewhat unfamiliar & Somewhat unfamiliar & Chores, remind about deadlines\\
2 & {F02} & - & Math & Neutral & Somewhat unfamiliar & Chores\\
3 & {F03} & - & Food Systems & Somewhat unfamiliar & Unfamiliar & Cooking\\
4 & {F04} & - & Applied Math & Unfamiliar & Unfamiliar & Cooking, driving\\
5 & {F05} & - & Math & Somewhat unfamiliar & Unfamiliar & Manage files\\
6 & {F06} & - & Music & Neutral & Unfamiliar & Grab packages, sort folders\\
7 & {F07} & - & Applied Math & Somewhat familiar & Unfamiliar & Fetch items\\
8 & {F08} & - & Computer Science & Somewhat unfamiliar & Somewhat unfamiliar & Cooking, driving, grab packages\\
9 & {F09} & - & Statistics & Somewhat unfamiliar & Unfamiliar & Chores, teaching\\
10 & {F10} & - & Computer Science & Neutral & Somewhat unfamiliar & Assemble car components\\
11 & - & {E01} & Music & Unfamiliar & Unfamiliar & Grab packages\\
12 & - & {E02} & Electrical Engineering & Somewhat unfamiliar & Unfamiliar & Chores\\
13 & - & {E03} & Statistics & Unfamiliar & Unfamiliar & Fetch items\\
14 & - & {E04} & Business & Somewhat unfamiliar & Somewhat unfamiliar & Chores\\
15 & - & {E05} & Computer Science & Somewhat unfamiliar & Unfamiliar & Fold clothes\\
16 & - & {E06} & Electrical Engineering & Neutral & Somewhat unfamiliar & Cooking\\
17 & - & {E07} & Education & Somewhat unfamiliar & Unfamiliar & Teaching\\
18 & - & {E08} & Informatics & Somewhat unfamiliar & Unfamiliar & Grab packages\\
19 & - & {E09} & Genetics & Neutral & Unfamiliar & Cleaning\\
20 & - & {E10} & Computer Science & Neutral & Neutral & Sort folders\\
21 & - & {E11} & Informatics & Somewhat unfamiliar & Somewhat unfamiliar & Grab packages\\
22 & - & {E12} & Math & Unfamiliar & Unfamiliar & Teaching\\
23 & - & {E13} & Robotics & Familiar & Neutral & Fold clothes\\
24 & - & {E14} & Mechanical Engineering & Familiar & Somewhat familiar & Factory inspection, fetch items\\
\bottomrule
\end{tabular}
\caption{Participant demographics in formative and evaluation studies. There were a total of 24 participants overall. \textit{f} denotes participant IDs for the formative study, and \textit{e} denotes participant IDs for the evaluation study.}
\Description{Participant demographics in formative and evaluation study. There were 24 participants overall (#), 10 for the formative study (f denotes the participant IDs) and 14 for the evaluation study. We recorded the area of expertise, familiarity with robotics (5-point Likert scale), familiarity with demonstration collection (5-point Likert scale), and preferred tasks for the robot.}
\end{table*}

\end{document}